\documentclass[
    a4paper, 
    twocolumn, 
    superscriptaddress, 
    amsmath, 
    amssymb, 
    amsfonts, 
    floatfix
]{revtex4-2}

\usepackage[utf8]{inputenc}
\usepackage[english]{babel}
\usepackage{amsmath,amsfonts,amssymb}
\usepackage{mathpazo}
\usepackage[scaled]{helvet}
\usepackage[T1]{fontenc}
\usepackage{url}
\usepackage{hyperref}
\hypersetup{colorlinks=true, allcolors=blue}

\usepackage{graphicx,xcolor}
\usepackage{microtype}
\usepackage{graphicx}

\begin{document}

\title{Toward integrated tantalum pentoxide optical parametric oscillators}
\date{June 5, 2023}

\author{Maximilian Timmerkamp}
\email{m.timmerkamp@uni-muenster.de}
\affiliation{University of Münster, Institute of Applied Physics, Corrensstr. 2, 48149 Münster, Germany}
\author{Niklas M. Lüpken}
\affiliation{University of Münster, Institute of Applied Physics, Corrensstr. 2, 48149 Münster, Germany}
\author{Shqiprim Adrian Abazi}
\affiliation{University of Münster, Institute of Physics, Heisenbergstr. 11, 48149 Münster, Germany}
\affiliation{Center for NanoTechnology (CeNTech), Heisenbergstr. 11, 48149 Münster, Germany}
\affiliation{Center for Soft Nanoscience (SoN), Busso-Peus-Str. 10, 48149 Münster, Germany}
\author{Julian Rasmus Bankwitz}
\affiliation{University of Münster, Institute of Physics, Heisenbergstr. 11, 48149 Münster, Germany}
\affiliation{Center for NanoTechnology (CeNTech), Heisenbergstr. 11, 48149 Münster, Germany}
\affiliation{Center for Soft Nanoscience (SoN), Busso-Peus-Str. 10, 48149 Münster, Germany}
\author{Carsten Schuck}
\affiliation{University of Münster, Institute of Physics, Heisenbergstr. 11, 48149 Münster, Germany}
\affiliation{Center for NanoTechnology (CeNTech), Heisenbergstr. 11, 48149 Münster, Germany}
\affiliation{Center for Soft Nanoscience (SoN), Busso-Peus-Str. 10, 48149 Münster, Germany}
\author{Carsten Fallnich}
\affiliation{University of Münster, Institute of Applied Physics, Corrensstr. 2, 48149 Münster, Germany}

%%%%%%%%%%%%%%%%%%% abstract %%%%%%%%%%%%%%%%
\begin{abstract}
We present a hybrid waveguide-fiber optical parametric oscillator (OPO) exploiting degenerate four-wave mixing in tantalum pentoxide. The OPO, pumped with ultrashort pulses at 1.55\,µm wavelength, generated tunable idler pulses with up to 4.1\,pJ energy tunable between 1.63\,µm and 1.68\,µm center wavelength. An upper bound for the total tolerable cavity loss of 32\,dB was found, rendering a chip-integrated OPO feasible as a compact and robust light source.
\end{abstract}

\maketitle

%%%%%%%%%%%%%%%%%%%%%%%%%%  body  %%%%%%%%%%%%%%%%%%%%%%%%%%
\section{Introduction}
Optical parametric oscillators (OPOs) are versatile light sources for a variety of applications such as nonlinear microscopy\cite{Gottschall2015, Brinkmann2019}, spectroscopy\cite{Adler2010}, and generation of squeezed light\cite{Tanimura2006}.
In recent years, fiber-based OPOs have been developed, exploiting four-wave mixing (FWM) for optical parametric gain and providing compact and robust light sources\cite{Brinkmann2019, Gottschall2015}. 
To reduce the size and the required pump power as well as to simplify thermal stabilization of the light source, fibers can be replaced by integrated optical waveguides.

Waveguide-based OPOs (WOPOs) have been demonstrated as hybrid waveguide--fiber devices using silicon\cite{Wang2015i,Kuyken2013} or silicon nitride\cite{Lupken2021} (Si$_3$N$_4$) waveguides to provide parametric gain and fibers to build the feedback cavity. Although silicon\cite{Bristow2007} provides a 100 times higher nonlinearity than Si$_3$N$_4$\cite{Ikeda2008}, it suffers from significant linear and nonlinear losses, which render integration of the cavity on a chip unfeasible at repetition rates below a few 100\,MHz. In contrast, Si$_3$N$_4$ allows for ultra-low loss propagation\cite{Liu2021}, such that integration of the cavity has been proposed to be feasible at repetition rates as low as 66\,MHz\cite{Lupken2021}.

Recently, tantalum pentoxide (Ta$_2$O$_5$) gained attraction in integrated nonlinear photonics\cite{Jung2021, Woods2020, Black2021, Chen2011}, offering optical properties competitive with Si$_3$N$_4$. Although the fabrication of Ta$_2$O$_5$ waveguides is less mature, both materials share the potential for ultra-low loss propagation\cite{Belt2017, Jung2021}. Ta$_2$O$_5$ shows an about three times higher nonlinear refractive index of approximately $6\cdot10^{-19}$\,m$^{2}$\,W$^{-1}$\cite{Jung2021, Tai2004, Ikeda2008}, which accomplishes decreased pump energies, and a ten times smaller thermo-optic coefficient of approximately $6\cdot10^{-6}$\,K$^{-1}$\cite{Wu2019, Jung2021}, allowing for light sources with higher thermal stability. Moreover, Ta$_2$O$_5$ allows for sustaining at least double the optical power compared to Si$_3$N$_4$\cite{Ahluwalia2009} and offers more design freedom for satisfying phase-matching conditions in nonlinear optical frequency mixing processes because of its low internal stress that facilitates the deposition of thicker films\cite{Jung2021}. As such, Ta$_2$O$_5$ waveguides have been used to generate supercontinua\cite{Woods2020}, Kerr-combs with micro-ring resonators\cite{Black2021}, and parametric amplification\cite{Chen2011}, while also being investigated for integrated quantum optics\cite{Splitthoff2020}.

In this work, we present a proof-of-concept WOPO exploiting degenerate FWM in Ta$_2$O$_5$ waveguides including a feedback cavity implemented with an optical fiber. The WOPO was pumped with ultrashort pulses and its output wavelength was dispersively tuned by changing the cavity length. The oscillation threshold and the maximum acceptable cavity loss were investigated to estimate the feasibility for integration of the whole cavity on a chip.

\section{Experimental setup} \label{sec:setup}
\begin{figure}
    \centering
    \includegraphics[width=1\linewidth]{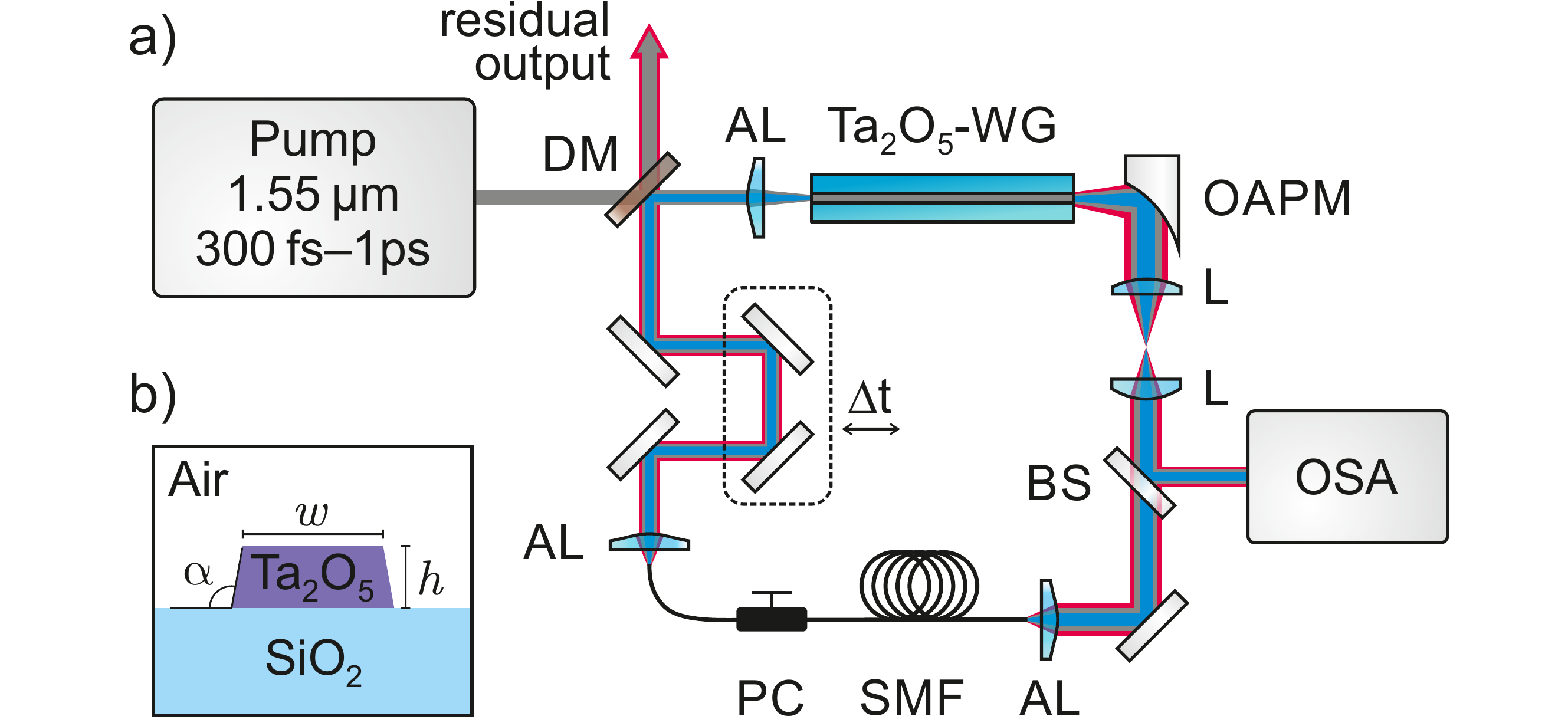}
    \caption{\label{fig:setup}(a) Schematic experimental setup. DM:~dichroic mirror, WG:~waveguide, AL:~aspheric lens, OAPM:~off-axis parabolic mirror, L:~lens, BS:~beam splitter, OSA:~optical spectrum analyzer, SMF:~single mode fiber, PC:~polarization controller. (b) Schematic waveguide geometry with sidewall angle $\alpha\approx95^{\circ}$.}
\end{figure}

The experimental setup of the WOPO (see Fig.~\ref{fig:setup}(a)) comprised a Ta$_2$O$_5$ waveguide as the parametric gain medium and a fiber to build the feedback ring cavity. The WOPO was pumped with pulses at 1.55\,µm center wavelength with a repetition rate of 80\,MHz. The duration of the pump pulses was adjusted with a Fourier-filter (not shown) between 300 and 1000\,fs. The pump pulses were coupled into an air-cladded waveguide (H$\times$W$\times$L$\,=\,$0.6\,µm$\times$1.7\,µm$\times$10\,mm, schematically shown in Fig.~\ref{fig:setup}(b)) with an aspheric lens (AL, $f=1.873$\,mm, $\text{NA}=0.85$), exciting the fundamental TE mode. 

The waveguides were fabricated from a 0.6\,µm thin Ta$_2$O$_5$ film on thermally grown silicon dioxide on silicon to achieve strong mode confinement due to high refractive index contrast. Waveguide layouts were patterned in a resist mask using a 100\,kV electron-beam lithography system and transferred into the Ta$_2$O$_5$ thin film via subsequent reactive-ion etching in fluorine chemistry (Ar, CF$_4$ and CHF$_3$). 
The desired group velocity dispersion (GVD) was achieved with high reproducibility in fully etched waveguides of suitable geometry and the facets were cleaved after inscribing a kerf on the backside of the handle wafer with a dicing system.

The output from the waveguide was collected with an off-axis parabolic mirror (OAPM, $f_\text{eff}=6.35$\,mm) to minimize chromatic aberrations. The total transmission (including input AL and output OAPM) at the pump wavelength was 2.6\% due to several contributions: mode mismatch ($\ge$7\,dB loss) caused by residual higher-order transverse modes in the pump beam, absorption of the aspheric lens (1.1\,dB loss), truncation loss at the OAPM (3.0\,dB), Fresnel reflections at the facets (each 0.4\,dB loss), and $\le$4\,dB/cm propagation loss along the waveguide resulting mainly from fabrication imperfections, including stitching of writing fields.

Inside the waveguide, signal and idler sidebands were generated via spontaneous degenerate FWM, of which the signal sideband was used to implement a feedback for oscillation. The beam coming from the OAPM was adjusted in diameter with a telescope for coupling into a single-mode fiber SMF-28. In order to match the repetition rates of pump and WOPO for harmonic pumping as well as to implement an amount of group delay dispersion (GDD) similar to a chip-integrated feedback path (-0.7\,ps$^2$ to 3.3\,ps$^2$, depending on waveguide geometry), a feedback fiber length of 38\,m was chosen, introducing a GDD of -0.59\,ps$^2$ at the signal wavelength.
A fiber polarization controller was used to adjust the polarization of the fed-back signal pulse to match the TE-polarization of the pump. The signal pulses were stretched by the fiber's GVD of -16\,ps$^2$\,km$^{-1}$, enabling dispersive tuning of the signal wavelength due to temporal gain-narrowing\cite{Gottschall2015} by adjustment of the temporal overlap of signal and pump pulse via a free-space delay line ($\Delta t$ in Fig.~\ref{fig:setup}).
Then, the signal beam was spatially overlapped with the pump beam at the dichroic mirror (DM) and coupled into the waveguide to provide a seed for stimulated FWM, leading to parametric oscillation when the gain was larger than the cavity loss.
Since the signal pulse provided feedback, the idler as well as the residual pump pulse could have been completely extracted from the cavity with the addition of a second dichroic mirror. 
However, to analyze the whole spectrum instead of only the pump and idler waves, an uncoated glass substrate (BS) was used to extract a fraction of the intra-cavity field for detection with an optical spectrum analyzer (OSA). Unless noted otherwise, all pump energies in the following parts refer to (measurable) waveguide-external energies incident on the input aspheric lens in front of the waveguide.
Output energies of the signal and idler waves refer to cavity-internal values, externally measured and accordingly rescaled.

\section{Operation characteristics}

\begin{figure}
    \centering
    \includegraphics[width=1\linewidth]{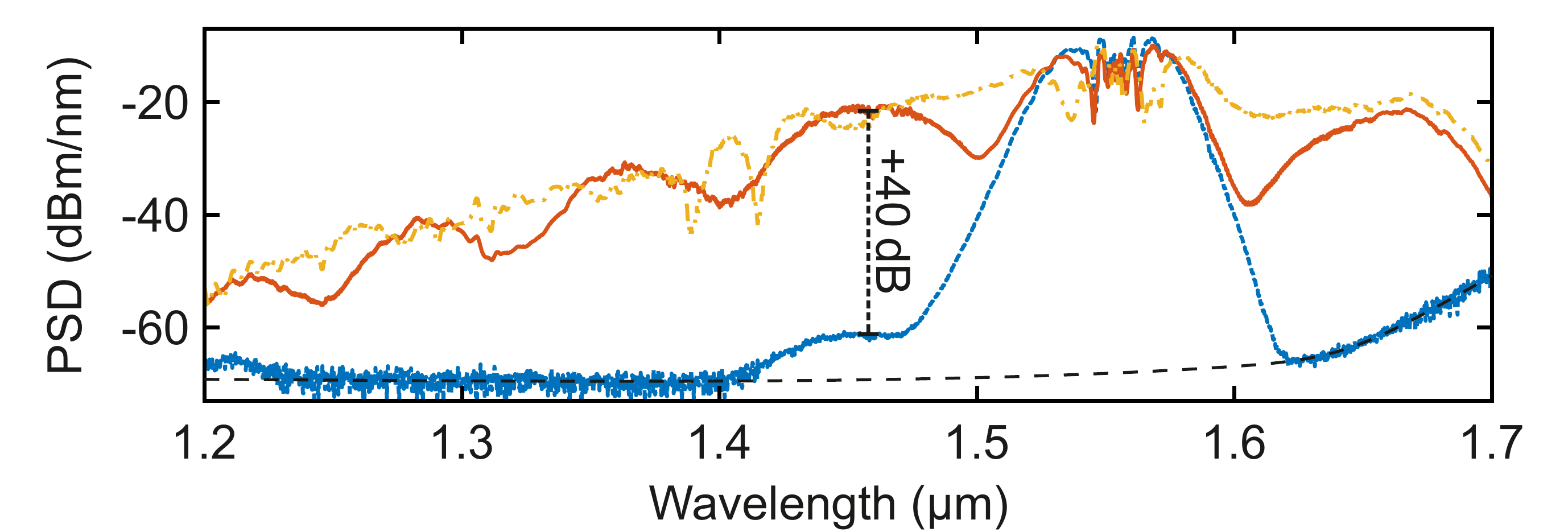}
    \caption{\label{fig:spectra}Power spectral density (PSD) of spontaneous FWM (dashed, blue) and the output of the oscillating WOPO (solid, red) for a 38\,m long feedback fiber and a pump pulse with 500\,fs duration and 2.2\,nJ energy. Additionally, a spectrum for 2.75\,nJ pump energy (dash-dotted, yellow) and the OSA's detection background (black, dashed) is included.}
\end{figure}

Pumping the WOPO with pulses of 500\,fs duration and 2.2\,nJ energy, an amplification of 40\,dB was achieved at the signal wavelength of 1.46\,µm after feedback was implemented. In addition to the first-order signal and idler peaks at 1.46\,µm and 1.67\,µm wavelength, respectively, the output spectra showed additional peaks due to cascaded FWM (e.g., at 1.37\,µm, see Fig.~\ref{fig:spectra}). Although the duration of the idler pulse was not measured due to insufficient power output at the OSA port, a potential Fourier-limited pulse duration of (130$\pm$10)\,fs was calculated from the (first-order) idler spectrum. The actual pulse duration was expected to be approximately 20\% longer, which was estimated by comparing the actual and the Fourier-limited idler pulse durations of simulations of the WOPO by solving the corresponding nonlinear Schrödinger equation.

\begin{figure}
    \centering
    \includegraphics[width=1\linewidth]{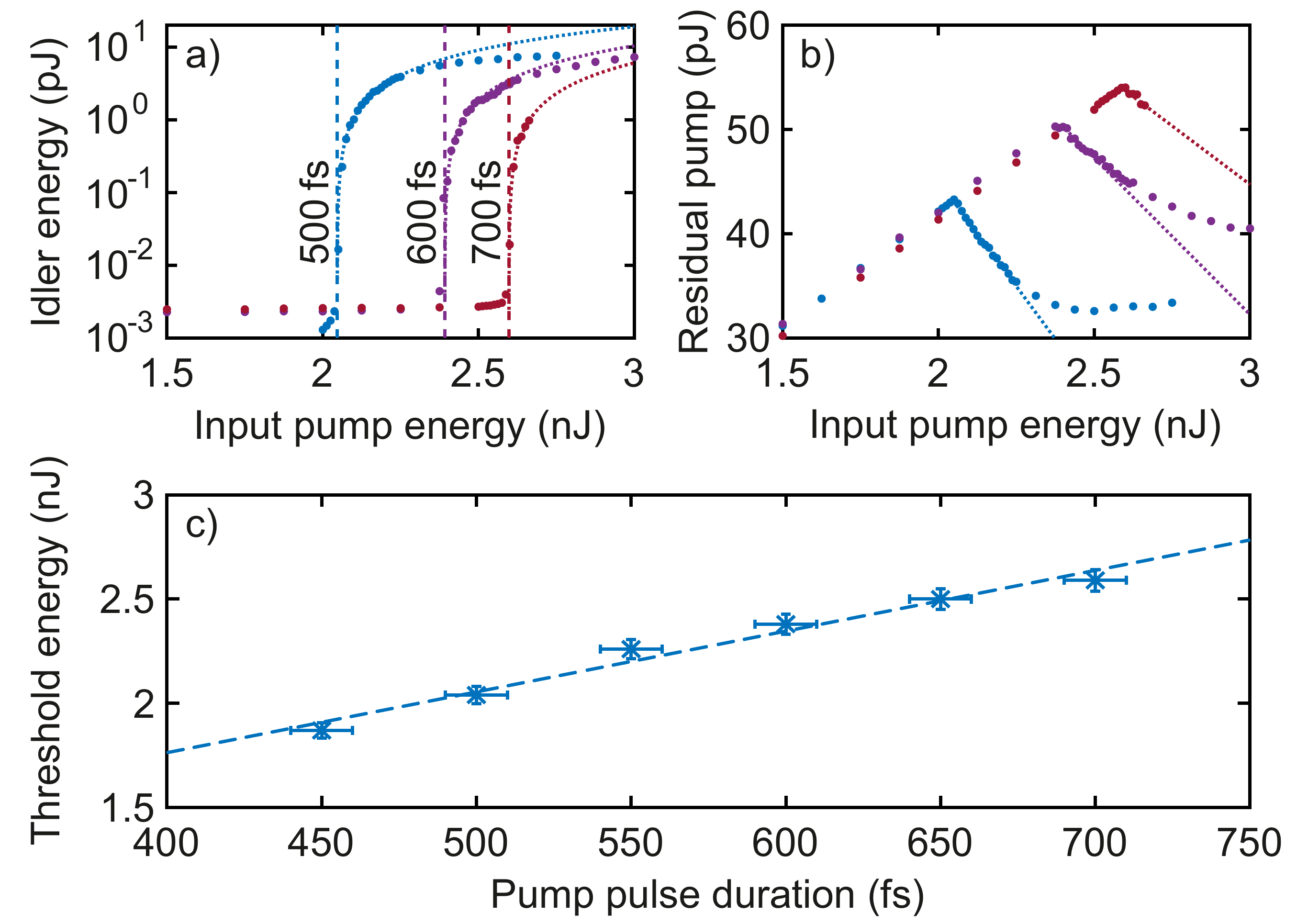}
    \caption{\label{fig:osc_threshold}(a) Energy in the idler sideband (dots) versus external pump pulse energy. Linear functions (dotted) were fitted to the data. The roots of these functions mark the threshold values (dashed lines). (b) Residual pump energy at waveguide output (dots) with pulse parameters corresponding to (a). (c) Oscillation threshold (crosses) versus pump pulse duration with linear fit (dashed).}
\end{figure}

In order to determine the oscillation threshold, the output spectrum was measured for increasing pump pulse energies, showing a steep increase of energy in the sidebands once the threshold was exceeded. Oscillation was achieved at a pump energy of approximately 2\,nJ (i.e. $\le$280\,pJ waveguide-internally) for a pump pulse duration of 500\,fs (see Fig.~\ref{fig:osc_threshold}(a)). The threshold values were identified with the roots of linear least-square fits near the threshold. At pump pulse energies significantly higher than the threshold, saturation was observed for the idler energy as well as the pump depletion (see Fig.~\ref{fig:osc_threshold}(a) and \ref{fig:osc_threshold}(b)).
In the saturated regime, the clear FWM peaks gradually merged and vanished (compare dash-dotted line in Fig.~\ref{fig:spectra}), indicating transition into supercontinuum generation (SCG) induced by modulation instability (MI) seeded by the fed-back signal pulse.
The average waveguide-external slope efficiency just above the threshold was $(1.9\pm 0.8)\%$ and the maximum waveguide-external conversion efficiency of -27.3\,dB (0.19\%) was observed at 2.25\,nJ pump energy, when the idler sideband showed a total energy of 4.1\,pJ. Considering the combined coupling and collection efficiency of about 10\%, the waveguide-internal slope and conversion efficiency amounted to -17.3\,dB (1.9\%) and $(19\pm 8)\%$, respectively. In comparison to other works on WOPOs\cite{Kuyken2013,Wang2015i,Lupken2021}, the efficiency values observed here were comparable but at the lower end due to present fabrication imperfections. However, the residual pump at the output in Fig.~\ref{fig:osc_threshold}(b) showed up to about 20\% pump depletion, such that up to 9\% conversion efficiency to the idler can be expected with improved coupling and collection efficiencies.

The operation of the WOPO was investigated for pump pulse durations between 450\,fs and 700\,fs as lower pulse durations led to soliton-driven SCG, and insufficient pump energy inhibited oscillation for higher pulse durations.
The oscillation threshold increased linearly with the pump pulse duration (see Fig.~\ref{fig:osc_threshold}(c)), indicating an approximately constant peak power value at threshold.
This observation is in compliance with the proportionality between pump peak power and (maximum) parametric FWM gain, which, at threshold, exactly compensates the constant cavity loss.

\section{Wavelength tunability}
The center wavelength of the idler (signal) sideband was dispersively tuned by changing the cavity length via the free-space delay line, covering a tuning range of 50\,nm (40\,nm). 
To operate the WOPO outside the SCG regime over the whole tuning range, it was pumped by pulses of 2.2\,nJ energy and 500\,fs duration.

\begin{figure}
    \centering
    \includegraphics[width=1\linewidth]{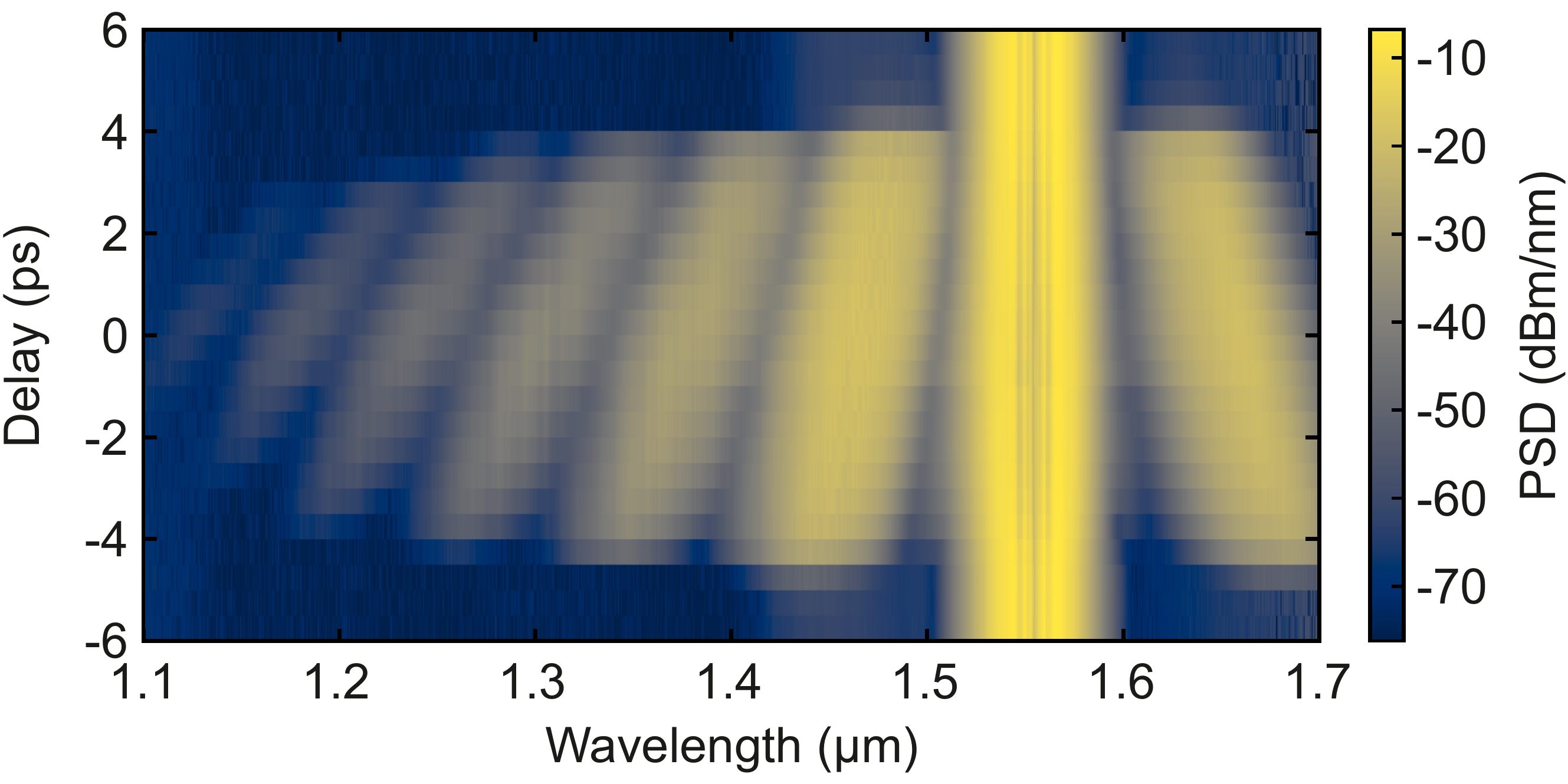}
    \caption{\label{fig:tuning}Power spectral density (PSD) of the WOPO output for different temporal delays between pump and fed-back signal pulses using pump pulses with 2.2\,nJ energy and 500\,fs duration.}
\end{figure}
 
By changing the relative delay between pump and feedback pulses by 8\,ps, the center wavelength of the idler and signal waves was tuned from 1.63\,µm to 1.68\,µm and from 1.44\,µm to 1.48\,µm, respectively (see Fig.~\ref{fig:tuning}). Since the FWM gain, indicated by the spontaneous FWM in Fig.~\ref{fig:spectra}, must exceed the cavity round-trip loss (98.7\%) for oscillation, the tuning range was smaller than the potential gain bandwidth. Furthermore, the tuning range was limited by chromatic aberrations caused by the lenses, and the bandwidth of the effective wave plate induced by the polarization controller, which was optimized only once for 1.46\,µm signal wavelength.
In the future, lower cavity round-trip loss and a polarization-maintaining feedback path should allow for an increased accessible tuning range, while using a 2\,µm wide waveguide should extend the potential wavelength tuning range even to 250\,nm (i.e., 46\,THz bandwidth from 1.15\,µm to 1.4\,µm wavelength) for the signal wave or 640\,nm (i.e., 46\,THz bandwidth from 1.74\,µm to 2.38\,µm wavelength) for the idler wave.

\section{Acceptable cavity loss} \label{sec:loss}
Additional loss was introduced into the cavity in order to estimate the feasibility of integrating the cavity on a chip with respect to propagation loss. Additionally, this information can be used to determine the fraction of the signal pulse that could be extracted from the cavity with an additional output coupler. By placing a combination of a half-wave plate and a polarizing beam splitter between the delay stage and the dichroic mirror, up to 30\,dB additional loss was added to the cavity. Without additional loss the cavity showed a round-trip transmission of 1.3\% (-19\,dB), which was estimated from transmission values measured at the pump wavelength: 2.6\% through the waveguide, 65\% through the fiber, and 79\% through all other components.
The pump pulse duration (500\,fs) and the delay stage were fixed, such that the threshold was minimized. Then, the oscillation threshold was determined by measuring the energy of the signal sideband for a set of additional losses from 0\,dB to 30\,dB, corresponding to -19\,dB to -49\,dB feedback.

\begin{figure}
    \centering
    \includegraphics[width=1\linewidth]{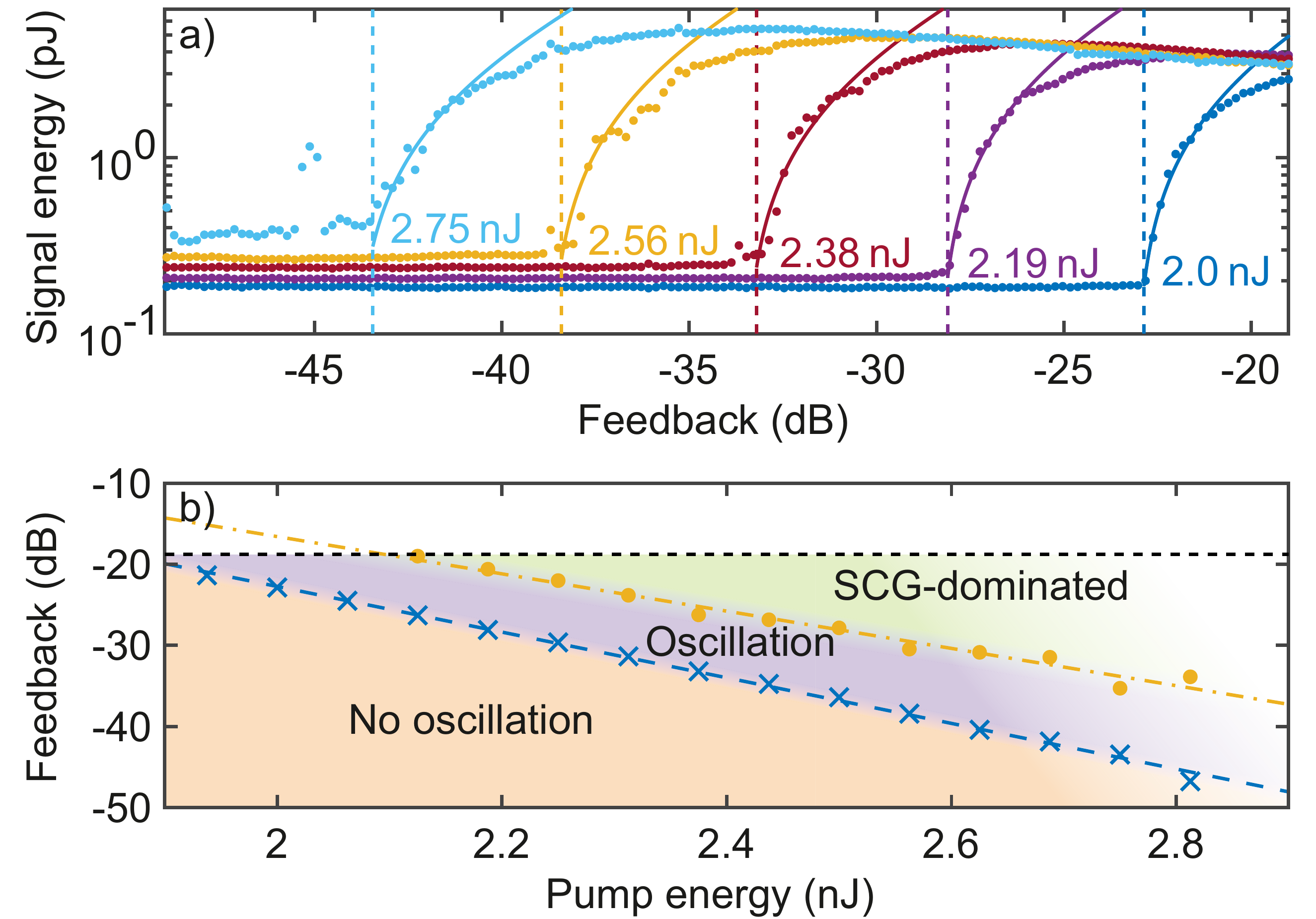}
    \caption{\label{fig:feedback}(a) Energy in signal sideband versus feedback of signal pulse for different pump pulse energies (dots). Additional linear fits (solid) of the rising flank and the feedback threshold  for oscillation (dashed) are plotted. (b) Critical feedback for oscillation (blue crosses) with linear fit (blue, dashed) and feedback at which the maximum signal energy occurred (yellow dots) with linear fit (dash-dotted). Also denoted are the different operation regimes of the WOPO.}
\end{figure}

The experimental data in Fig.~\ref{fig:feedback}(a) shows that pumping at a higher pump pulse energy allowed the WOPO to run with lower feedback. This behaviour is expected since pumping with a higher energy provides higher gain that can compensate the additional loss. For the same reason, the background signal, caused by spontaneous FWM, also increased. Therefore, to characterize the WOPO loss tolerance, the critical feedback (i.e., minimum feedback for oscillation, see Fig.~\ref{fig:feedback}(b)) for each pump energy was identified with the intersection between the background and linear least-square fits to the experimental data.

From Fig.~\ref{fig:feedback}(b), it is apparent that the critical feedback exponentially decreased for increasing pump energies. This observation can be explained by the fact that the gain increases exponentially with increasing pump energy, compensating the reduced feedback. When the feedback was increased beyond the critical value, the energy in the signal sideband first exhibited a maximum and then decreased due to excess feedback causing transition to MI-induced SCG. Hence, the feedback at which the maximum signal energy was reached (yellow dots in Fig.~\ref{fig:feedback}(b)) can be considered as an upper limit for stable operation.

Although higher pump energies reduced the critical feedback, fluctuations of the signal energy were observed for pump energies larger than 2.38\,nJ, indicating less stable operation. The reason for this unexpected observation is not yet fully understood, however, the measured fluctuations in the signal energy of the WOPO appear to be caused by the high parametric gain resulting in an increased sensitivity to slightly fluctuating round-trip transmission due to optomechanical influences and a limited achievable extinction ratio behind the fiber polarization controller. Therefore, for reliable, stable operation, a pump energy of less than 2.38\,nJ was used, corresponding to a minimum critical feedback of -32\,dB or an additional loss of 13\,dB, so that up to 95\% of the energy of the seed pulse could be extracted within the cavity.

\section{Feasibility of on-chip integration}
In order to implement the oscillator on a chip, both the nonlinear waveguide and the linear feedback path, whose minimum geometrical length of about 2\,m is related to the repetition rate of 80\,MHz and the effective refractive index of the TE-mode of about 1.9, must fit onto the chip. The feedback path should introduce a sufficient amount of GDD to allow for a tunable output while keeping the propagation loss low enough for oscillation.

In order to allow for dispersive tuning, the seed pulse has to be chirped by the feedback path. As already noted, the fiber introduced a GDD of -0.59\,ps$^2$ at 1.46\,µm signal wavelength, which allowed for dispersive tuning. Assuming the minimum chip-integrated cavity length of 2\,m, a suitable GDD at the signal wavelength in the range from -0.7\,ps$^2$ to 3.3\,ps$^2$ can be implemented on-chip by choosing a waveguide width between 1\,µm and 0.6\,µm, respectively.
Instead of using a free-space delay line, the WOPO could be tuned by changing the pump repetition rate or the pump center wavelength\cite{Brinkmann2019}.

As the minimum critical feedback required for oscillation was found to be -32\,dB at 2.38\,nJ pump energy, the total cavity loss at the seed wavelength must not exceed 32\,dB in order to operate the WOPO. Thus, given the minimum cavity length of 2\,m, a realistic upper boundary for propagation loss of 0.16\,dB/cm can be estimated. Since lower propagation losses have already been demonstrated in Ta$_2$O$_5$ waveguides\cite{Jung2021, Belt2017}, integration of the oscillator is principally possible concerning material selection and waveguide fabrication.

\section{Conclusion}
To sum up, a hybrid waveguide-fiber OPO (WOPO) exploiting four-wave mixing in a Ta$_2$O$_5$ waveguide was investigated in terms of its operation as well as feasibility of chip-integration. Reliable oscillation was achieved when the WOPO was pumped by pulses with 500\,fs duration and 2\,nJ waveguide-external energy, generating idler pulses with 4.1\,pJ waveguide-external energy and potentially down to 130\,fs duration. While the external conversion efficiency was only 0.19\%, the pump depletion indicated potential for up to 9\%. By adjusting the cavity length, the idler wavelength was dispersively tuned by 50\,nm from 1.63\,µm to 1.68\,µm, which can potentially be extended to up to 640\,nm (equiv. to 250\,nm for the signal wave) via waveguide dispersion engineering. By introducing loss to the cavity, the principal feasibility of a chip-integrated cavity, supporting 80\,MHz repetition rate, was confirmed by measurements, yielding an upper bound of 0.16\,dB/cm for the propagation loss, a value that has already been demonstrated in Ta$_2$O$_5$ waveguides\cite{Jung2021, Belt2017}. 

% \begin{backmatter}
% \bmsection{Funding}

% \bmsection{Acknowledgments}

{\par\medskip\noindent{\fontsize{9}{11}\bfseries {Disclosures}.\enspace}}
The authors declare no conflicts of interest.

{\par\medskip\noindent{\fontsize{9}{11}\bfseries {Data availability}.\enspace}} Data underlying the results presented in this paper are not publicly available at this time but may be obtained from the authors upon reasonable request.

% \end{backmatter}

%%%%%%%%%% If using BibTeX:
\bibliography{lit}

\end{document}